\documentstyle[12pt,epsf]{article}
\begin{document}
%
\title{Dielectric and Dilatometric Studies of Glass Transitions \\in Thin Polymer Films}
\author{Koji Fukao and Yoshihisa Miyamoto\\\normalsize\it
Faculty of Integrated Human Studies,
Kyoto University, Kyoto 606-8501,
Japan 
}
\date{(\small Received \today)}
\maketitle

\vspace*{-1cm}
\hspace*{1cm}\begin{minipage}{16cm}
\setlength{\baselineskip}{11pt}
\noindent{\small\bf Abstract.~}{\small
Dielectric relaxation and thermal expansion spectroscopy were made for 
thin polystyrene films in order to measure 
the temperature $T_{\alpha}$ corresponding to the peak in the loss 
component of susceptibility due to the $\alpha$-process and the
$\alpha$-relaxation time $\tau$ as functions of film thickness~$d$. 
While the glass transition temperature $T_{\rm g}$ decreases with 
decreasing film thickness,
$T_{\alpha}$ and $\tau$ were found to remain almost constant 
for $d>d_{\rm c}$ and 
decrease drastically for $d<d_{\rm c}$ for high temperatures. Here, 
$d_{\rm c}$ is a critical thickness.  Near the glass transition 
temperature, the thickness dependence of $T_{\alpha}$ and $\tau$ 
is more prominent.
The relation between the fragility index and non-exponentiallity is 
discussed for thin films of polystyrene.
}
\end{minipage}



\vspace*{0.7cm}\noindent{\bf 1. INTRODUCTION}\\

\vspace*{-0.1cm}\noindent Understanding the behavior of the characteristic length scale of the
dynamics of supercooled liquids near the glass transition is the most 
important problem to be solved in studies on glass transitions~\cite{Ediger1}.
%
Glass transitions in finite systems confined to
nanopores~[2-4] 
and thin films~[5-9] 
have recently attracted much attention, because such systems can be regarded 
as model systems for studying the length scale of glass transitions.  
In such systems, deviation from  
bulk properties is expected to appear if the system size is comparable
to the characteristic length scale. In particular, $T_{\rm g}$ of thin 
films have been measured using several experimental
techniques~[5-8] 
and a drastic reduction of $T_{\rm g}$ has been observed with
decreasing film thickness. 
However, the dynamics of the $\alpha$-process in thin polymer films 
has not yet been clarified in detail.

In previous papers~\cite{Fukao1}, we reported that $T_{\rm g}$ for thin
polystyrene films supported on glass substrate can be determined from
the temperature change of the electric capacitance 
and that the dynamics of the $\alpha$-process can be determined from the  
dielectric loss of the films. We were able to obtain the distinct thickness 
dependences of $T_{\rm g}$ and $T_{\alpha}$ in which the dielectric loss 
exhibits a peak value for a fixed frequency due to the $\alpha$-process.
In this paper, we investigate the dynamics of $\alpha$-process in thin 
polymer films, especially the thickness dependence of the dynamics for a
wider frequency range, including near the glass transition temperature, 
by applying 
dielectric relaxation (DES) and thermal expansion spectroscopy (TES).

\vspace*{0.6cm}\noindent{\bf 2. EXPERIMENTALS}\\

\vspace*{-0.1cm}\noindent Atactic polystyrene (a-PS) used in this study was 
purchased from the Aldrich 
Co., Ltd. ($M_{\rm w}$=
1.8$\times$10$^6$, $M_{\rm w}/M_{\rm n}$=1.03, where $M_{\rm w}$ and 
$M_{\rm n}$ are the weight average and the number average of the
molecular weights, respectively).
Thin films of a-PS with various thicknesses were prepared on an
Al-deposited slide glass using a 
spin-coat method from a toluene solution of a-PS. 
The thickness was controlled by changing the concentration of the solution. 
After annealing at 70$^{\circ}$C in the vacuum system for several days 
to remove solvents, Al was vacuum-deposited again to 
serve as an upper electrode. Heating cycles in which the temperature was 
changed between room temperature 
and 110$^{\circ}$C ($>$$T_{\rm g}$) were applied prior to  
the dielectric measurements to relax the as-spun films and obtain 
reproducible results. 
Dielectric measurements were done using an LCR 
meter (HP4284A) in the range of frequency of applied electric field,
$f_{\rm E}$, from 20 Hz to 1MHz during 
heating (cooling) processes in which the temperature was changed at a 
rate of 2K/min or 0.5K/min.

Thermal expansion spectroscopy is a new technique which has very
recently been introduced in studies on slow dynamics supercooled liquids 
by Bauer {\it et al}~\cite{Bauer1}. In this method, a sinusoidal temperature modulation,
$T(t)$=$\langle T\rangle$ + $T_{\omega}e^{i\omega t}$, is given 
to the sample, and then corresponding change in capacitance with the same
angular frequency as the temperature modulation has, 
$C'(t)$=$\langle C'\rangle$
+ $C'_{\omega}e^{i(\omega t+\delta )}$, is detected within a linear
response region. Here, $\omega$=2$\pi f_{\rm T}$ and $f_{\rm T}$ is the
frequency of temperature modulation.  Because this capacitance change is
directly connected with the volume change in thin films, the volume change 
can be obtained in accordance with the applied temperature modulation. 
Furthermore, the same sample can be used both for DES and TES without
changing measurement conditions.
In case of thin films of a-PS in which the area of 
the film surface remains constant with temperature change, the
temperature coefficient $\tilde\alpha$ of capacitance $C'$ and the
linear
thermal expansion coefficient, $\alpha_{\rm n}$, normal to the film surface  
satisfy the following relation:
$ \tilde\alpha\equiv -\frac{1}{C'(T_0)}\frac{dC'}{dT}\approx 2\alpha_{\rm n},$ 
where $C'(T)$ is the capacitance at temperature $T$ and $T_0$ is a
standard temperature. Using the amplitude of sinusoidal temperature modulation
$T_{\omega}$ and that of the response $C'_{\omega}$, $\tilde\alpha$ is given by the equation:
$\tilde\alpha$=$\frac{1}{C'(T_0)}\frac{C'_{\omega}}{T_{\omega}}$. In the
present measurements an average temperature $\langle T\rangle$ is
controlled to increase with a constant rate from 0.1 K/min to 0.5 K/min.  
The amplitude $T_{\omega}$ is set from 0.2 K to 0.6 K.
For capacitance measurement in TES the frequency of applied electric
field was chosen to be 100kHz to avoid the interference with dielectric 
relaxation.

\setcounter{section}{3}
\vspace*{0.6cm}\noindent{\bf 3. DIELECTRIC RELAXATION SPECTROSCOPY}\\


\noindent 
Dielectric relaxation spectroscopy for the frequency range 
$f_{\rm E}$=20Hz $\sim$ 1MHz reveals the dynamics of the
$\alpha$-process of a-PS at temperatures higher than the glass
transition temperature ($T_{\rm g}\sim 370$K) by ca. 20K $\sim$ 40K~\cite{Fukao1}. The results
obtained in this frequency and temperature range can be summarized as
follows: 1) The $d$ dependence of $T_{\rm g}$ is directly correlated to the 
distribution of relaxation times of the $\alpha$-process. 
2) The temperature at which dielectric loss is maximal  
due to the $\alpha$-process in the temperature domain and the
$\alpha$-relaxation time obtained by the frequency dependence of the 
dielectric 
loss remain constant down to the critical thickness $d_{\rm c}$, while
below $d_{\rm c}$ they decreases drastically with decreasing thickness.    
These results are shown  in Figs.2-4 with those of TES, where 
the observed points by
dielectric relaxation spectroscopy are plotted with the label ``DES''.

\vspace{0.6cm}\noindent{\bf 4. THERMAL EXPANSION SPECTROSCOPY}\\

\noindent 
Figure 1 shows the temperature change in both real and imaginary parts of 
dynamical thermal expansion coefficient $\alpha_{\rm n}$ at a frequency
$f_{\rm T}$=16.7mHz of temperature modulation. In case 
of $d$=362nm (bulk sample) the
loss peak exists at around 378K, while in case of films with $d$=18nm 
the peak temperature is shifted to lower temperature. At the same time,
the peak width clearly increases with decreasing film thickness. 
This $d$-dependence obtained by TES for low frequencies is quite similar 
to that obtained by DES for higher frequencies~\cite{Fukao1}. In bulk sample, 
the change in thermal expansion coefficient due to the glass transition occurs
in a narrow temperature range, while in very thin films of a-PS it
changes in a wider temperature range.
In Fig.1, the thermal expansion coefficient of glassy state,
$\alpha_{\rm g}$, is smaller than that in $d$=362nm at $f_{\rm T}$=
16.7mHz. 
However, it is observed that $\alpha_{\rm g}$ depends 

\vspace*{0.3cm}
\epsfxsize=16cm 
\centerline{
\hspace*{-0.6cm}\epsfbox{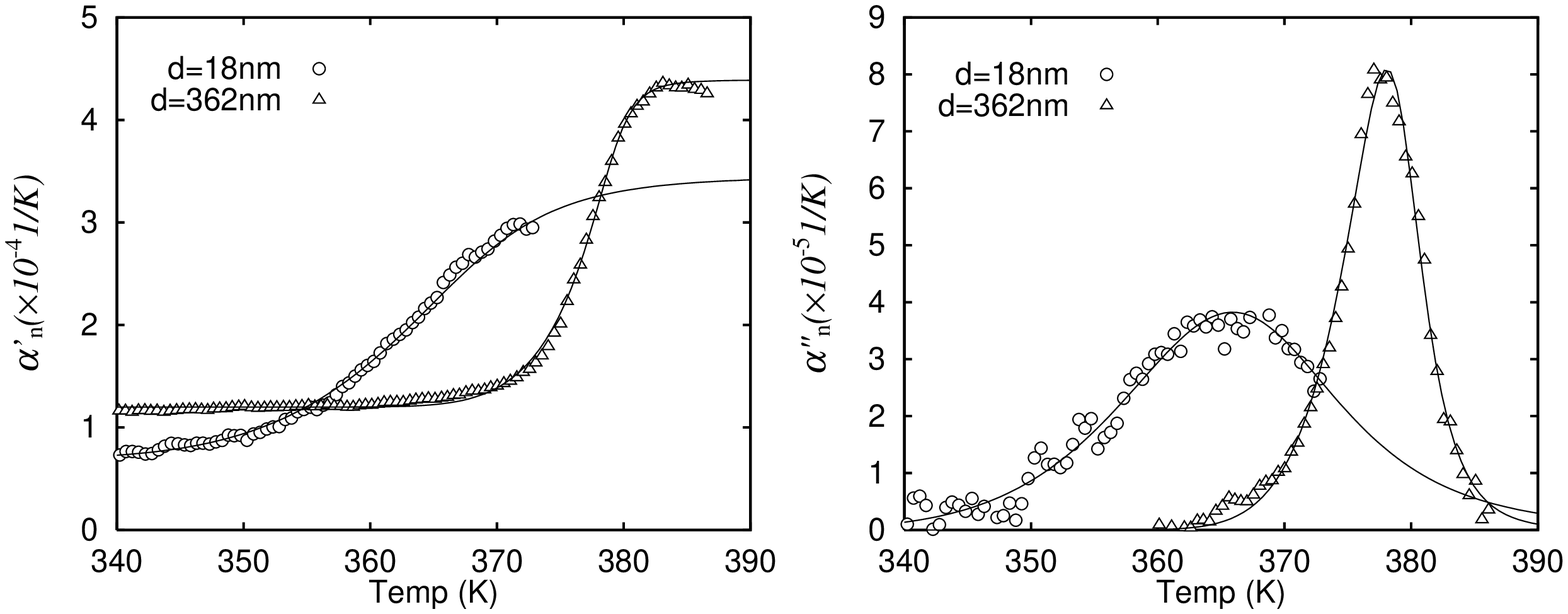}
}
{\setlength{\baselineskip}{9pt}\small
\noindent{\bf Figure 1.} Temperature dependence of complex linear thermal 
expansion coefficient $\alpha_{\rm n}$ for a-PS with film thickness 
18nm and 362nm ($f_{\rm T}$=1.67$\times$10$^{-2}$Hz).
The left figure shows the real part of $\alpha_{\rm n}$ and the right
one, the imaginary part. Solid lines are calculated by using the 
HN equation and the VFT equation.
}
\label{fig:fig1}
\vspace{0.2cm}

\vspace{0.5cm}
\noindent\begin{minipage}{8.75cm}
\epsfxsize=9cm 
\hspace*{-0.6cm}\epsfbox{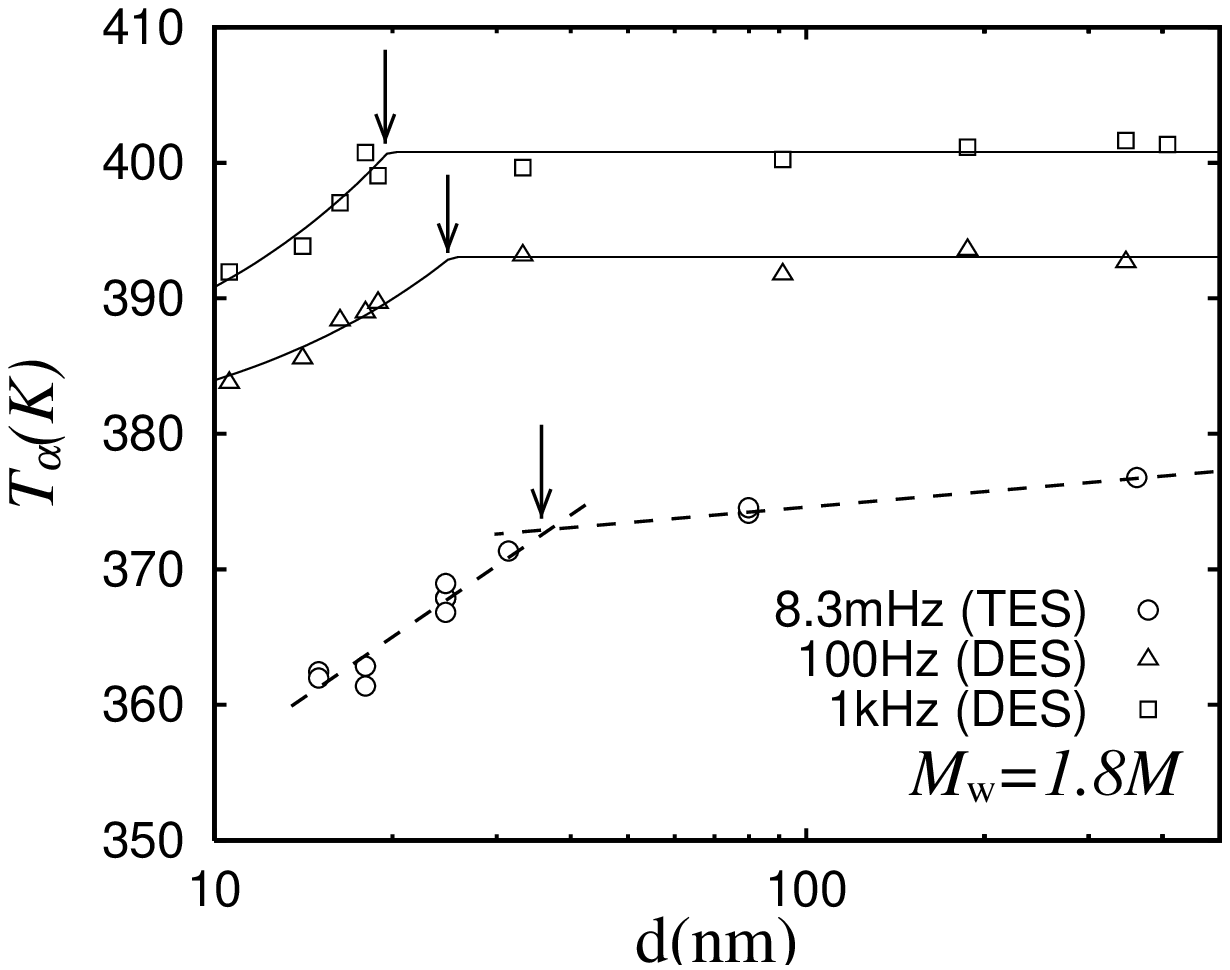}

{\setlength{\baselineskip}{9pt}
\small
{\bf Figure 2.} $d$ dependence of $T_{\alpha}$ 
of a-PS films obtained by DES and TES.
The arrows show critical thickness at which a drastic decrease in 
$T_{\alpha}$ begins. }

\vspace{0.0cm}
\end{minipage}
\hspace*{0.5cm}\vspace*{-0.5cm}\begin{minipage}{8.75cm}
\epsfxsize=9cm 
\hspace*{-0.6cm}\epsfbox{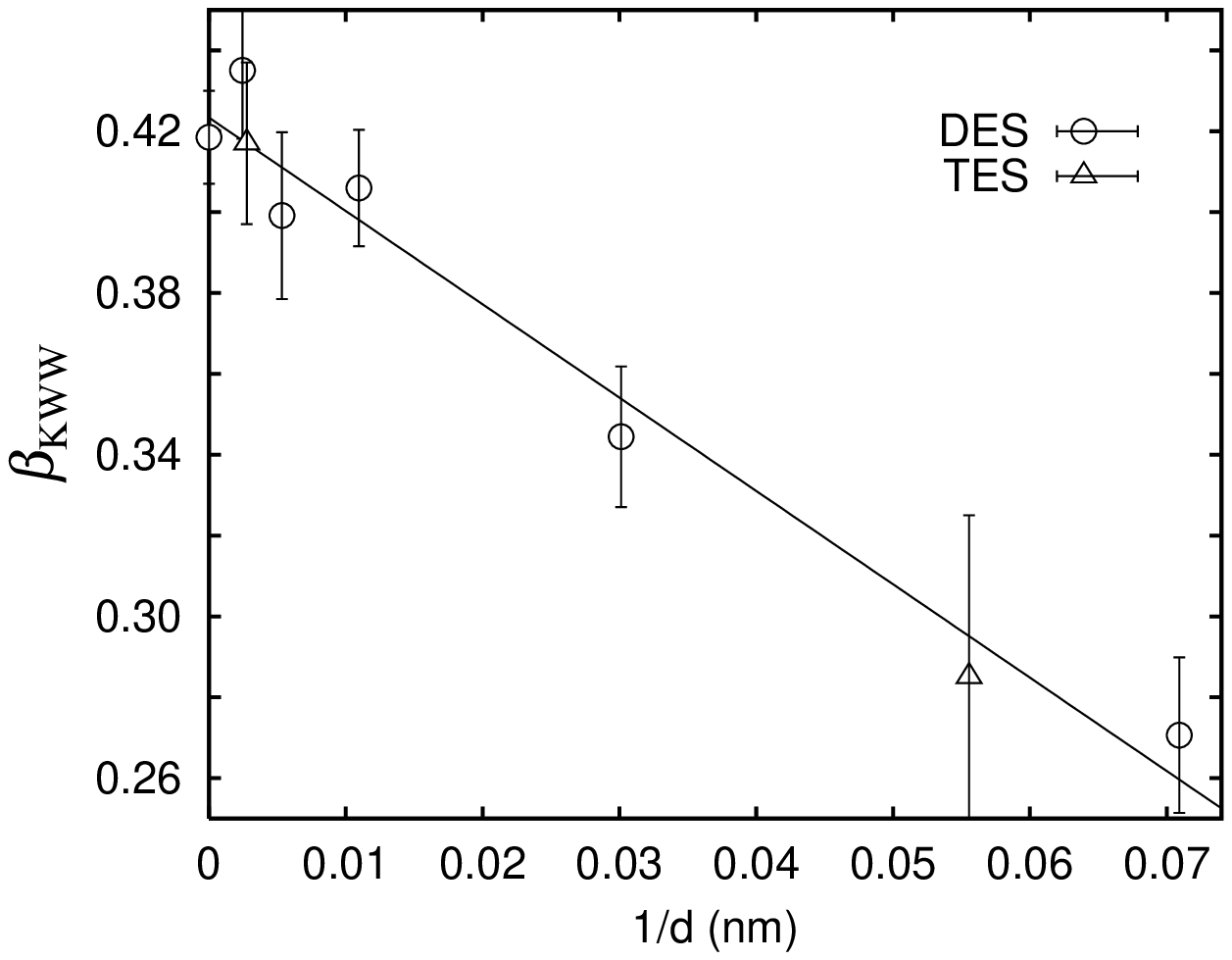}
\vspace{0.0cm}

{\setlength{\baselineskip}{9pt}
\small
{\bf Figure 3.} Thickness dependence of the KWW exponent
 $\protect\beta_{\mbox{\protect\tiny KWW}}$ 
obtained by DES ($\protect\circ$) and TES ($\protect\triangle$).
}

\vspace{0.4cm}
\end{minipage}


\vspace{1.0cm}\noindent 
on $f_{\rm T}$, i.e., $\alpha_{\rm g}$ 
increases with decreasing $f_{\rm T}$ and, as a result, 
$\alpha_{\rm g}$ for $d$=18nm is larger than that for $d$=362nm 
for the frequency $f_{\rm T}$ smaller than ca. 4mHz, which is consistent 
with the results obtained in previous works~\cite{Keddie1,Fukao1}.

\vspace*{0.6cm}\noindent{\bf 5. DYNAMICS OF $\alpha$-PROCESS IN a-PS THIN FILMS}\\

\noindent The $d$ dependence of $T_{\alpha}$ obtained by TES is 
shown in Fig.2, together
with that obtained by DES. 
The $d$ dependence of $T_{\alpha}$ observed by TES is qualitatively
similar to that by DES, although no detailed $d$ dependence of
$T_{\alpha}$ can so far be obtained by TES because of the limiting
number of the data with different thickness.  Nevertheless, it is
clearly seen that the critical thickness
$d_{\rm c}$ at which drastic decrease in $T_{\alpha}$ begins with
decreasing film thickness shifts to a larger value of $d$ as the
frequency of applied field decreases. A slight decrease can also be seen 
even in films with $d\approx$ 80nm for $f_{\rm T}$=8.3mHz by TES. Near
the glass transition temperature the $d$ dependence is more prominent 
compared with that at higher temperatures.

The dependence of $\alpha_{\rm n}$ on temperature can be reproduced by
assuming two empirical equations, as shown in Fig.1.
The solid curves in Fig.1 are calculated 
by using the Havriliak-Negami (HN) equation 
$\alpha_{\rm n}(\omega)$=$\alpha_{\rm
g}$+$\Delta\alpha/(1+(i\omega\tau )^{\alpha})^{\beta}$ and the 
Vogel-Fulcher-Tammann (VFT)
equation $\tau(T)$=$\tau_0\exp(U/(T-T_V))$, where
$\Delta\alpha$=$\alpha_{\rm l}-\alpha_{\rm g}$, $\alpha_{\rm l}$ is the
thermal expansion coefficient of the liquid state, $\alpha$ and $\beta$ 
are the shape parameters, $\tau_0$ and $U$ are constants, and $T_{\rm
V}$ is the Vogel temperature. Fitting the data in Fig.1 to 
the two equations, the shape parameters $\alpha$ and $\beta$ can be 
obtained: $\alpha$=0.46 and $\beta$=0.46 for $d$=18nm, and  
$\alpha$=0.85 and $\beta$=0.40 for $d$=362nm. The exponent
$\beta_{\mbox{\tiny KWW}}$ of the KWW relaxation function,
$\phi(t)=\exp(-(t/\tau)^{\beta_{\mbox{\tiny KWW}}})$, can be estimated
from the values of $\alpha$ and $\beta$ using the empirical relation 
$\beta_{\mbox{\tiny KWW}}\approx (\alpha\beta )^{1/1.23}$. As shown in Fig.3, 
it seems that the values of $\beta_{\mbox{\tiny KWW}}$
obtained by DES and TES fall on the same line, and that
$\beta_{\mbox{\tiny KWW}}$ becomes smaller as $d$
decreases. This result suggests that the distribution of the 
$\alpha$-relaxation times becomes broader as the thickness decreases.
This broadening may be due to the dynamical heterogeneity within 
thin films which is enhanced as the temperature approaches $T_{\rm g}$.  

We measured the temperature $T_{\alpha}$ for a given frequency $f_{\rm
E}$ and $f_{\rm T}$, and also the frequency $f_{\rm max}$ at which the
loss component has a maximum at a given temperature. The data observed
in both measurements by using both DES and TES can be summarized as
shown in Fig.4, which shows the dispersion map of the $\alpha$-process in thin a-PS
films for a frequency range from 10$^{-3}$Hz to 10$^4$Hz. The data 
measured for the films with the same or similar thickness are connected 
with each other using the VFT equation with same parameters in order to
obtain a temperature dependence of $\tau$ and $T_{\alpha}$ over a wider
temperature range. (See solid 
curves in Fig.4).  As suggested in Fig.2, 
the $d$ dependence of the $\alpha$-relaxation time changes with
frequency of applied field ($f_{\rm T}$ and $f_{\rm E}$). Near the glass
transition temperature, a shift of the $\alpha$-relaxation time with
change in $d$ in thin films is enhanced compared with that at higher
temperatures. For example, the frequency corresponding to the 
$\alpha$-relaxation time ranges over 3 decades at 375K, while it does
over a half decade at 400K. 

If we assume that there is a characteristic
length scale $\xi$ for the $\alpha$-process in this system, 
we can expect that the length scale $\xi$ increases
as the temperature approaches to $T_{\rm g}$. Accordingly, 
as $d$ decreases at a given temperature near $T_{\rm g}$, 
a deviation of  $\tau$   
from the values of bulk samples 
should begin at a larger value of $d$  than at higher temperatures, 
because the change of $\tau$ and $T_{\alpha}$ with $d$
becomes appreciable in case that $\xi$ reaches the thickness $d$.
This behavior is quite similar to that observed in
$\alpha$-dynamics in system confined in nano pores~\cite{Kremer1}. 

The frequency $f$ plotted in Fig.4 is related to 
the $\alpha$-relaxation time $\tau$ through the relation 
$2\pi f\tau=1$. Using this relation and the definition of
$T_{\rm g}$ that $\tau (T_{\rm g})$=10$^2$ sec, the data in Fig.4 can be 
replotted as shown in Fig.5, where so-called Angell plot is used.
In Fig.5, the fragility index $m$, which is defined by the equation 
$m$=$(\frac{d\log\tau(T)}{d(T_{\rm g}/T)})_{T=T_{\rm g}}$, can be 
estimated for a-PS films.  The values of $m$ for various $d$ values are  
as follows: $m$=86 for $d$=18nm, $m$=116 for 
$d$=30nm and 33nm, and  $m$=136 for $d=$362nm and 408nm. Hence, it is
found from the present measurements that the fragility index $m$
decreases with decreasing $d$, in other words, a-PS films become less
fragile as $d$ decreases. Because the exponent $\beta_{\mbox{\tiny
KWW}}$, which is a measure for non-exponentiality, decreases with 
decreasing $d$, as shown in Fig.3, the exponent $\beta_{\mbox{\tiny
KWW}}$ is an increasing function of the fragility index $m$.

On the other hand, it is well-known than in bulk supercooled liquids, 
the fragility index $m$ is strongly 
correlated with the exponent $\beta_{\mbox{\tiny KWW}}$; as $m$
increases, the exponent $\beta_{\mbox{\tiny KWW}}$
decreases, except a few groups of materials~\cite{Bohmer1}.
In case of thin films of a-PS, it is found that the general correlation
between $m$ and $\beta_{\mbox{\tiny KWW}}$ does not hold.

\noindent\begin{minipage}{8.75cm}
\epsfxsize=9cm 

\vspace*{-0.2cm}
\hspace*{-0.6cm}\epsfbox{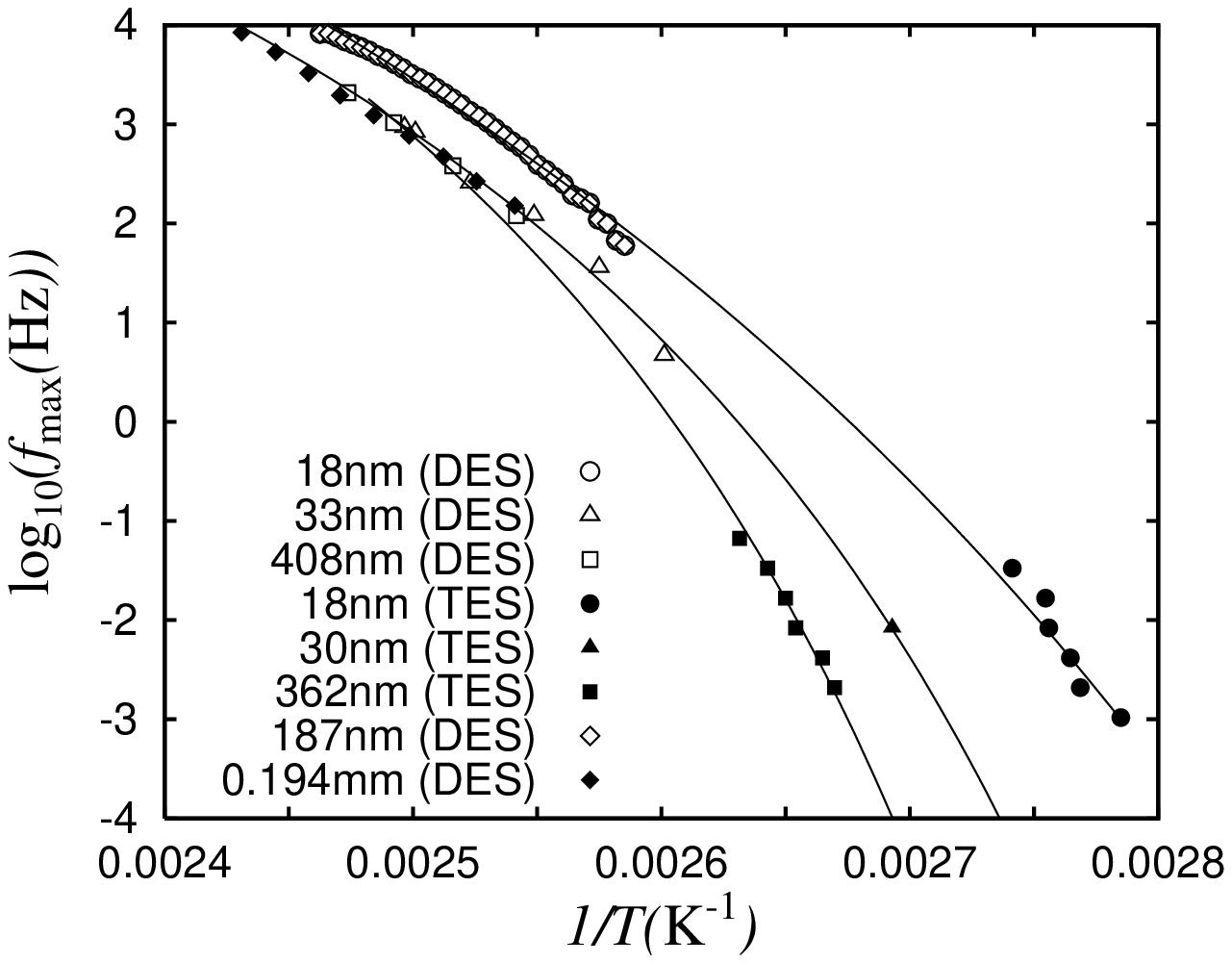}
\vspace{0.1cm}
{\setlength{\baselineskip}{9pt}
\small
{\bf Figure 4.} 
Dispersion map for thin films of a-PS obtained from the peak positions
of the loss component $\alpha''_{\rm n}$ or $\epsilon''$ 
for various film thicknesses. 
}

\vspace{0.0cm}
\end{minipage}
\hspace*{0.5cm}
\begin{minipage}{8.75cm}
\epsfxsize=9cm 
\hspace*{-0.6cm}\epsfbox{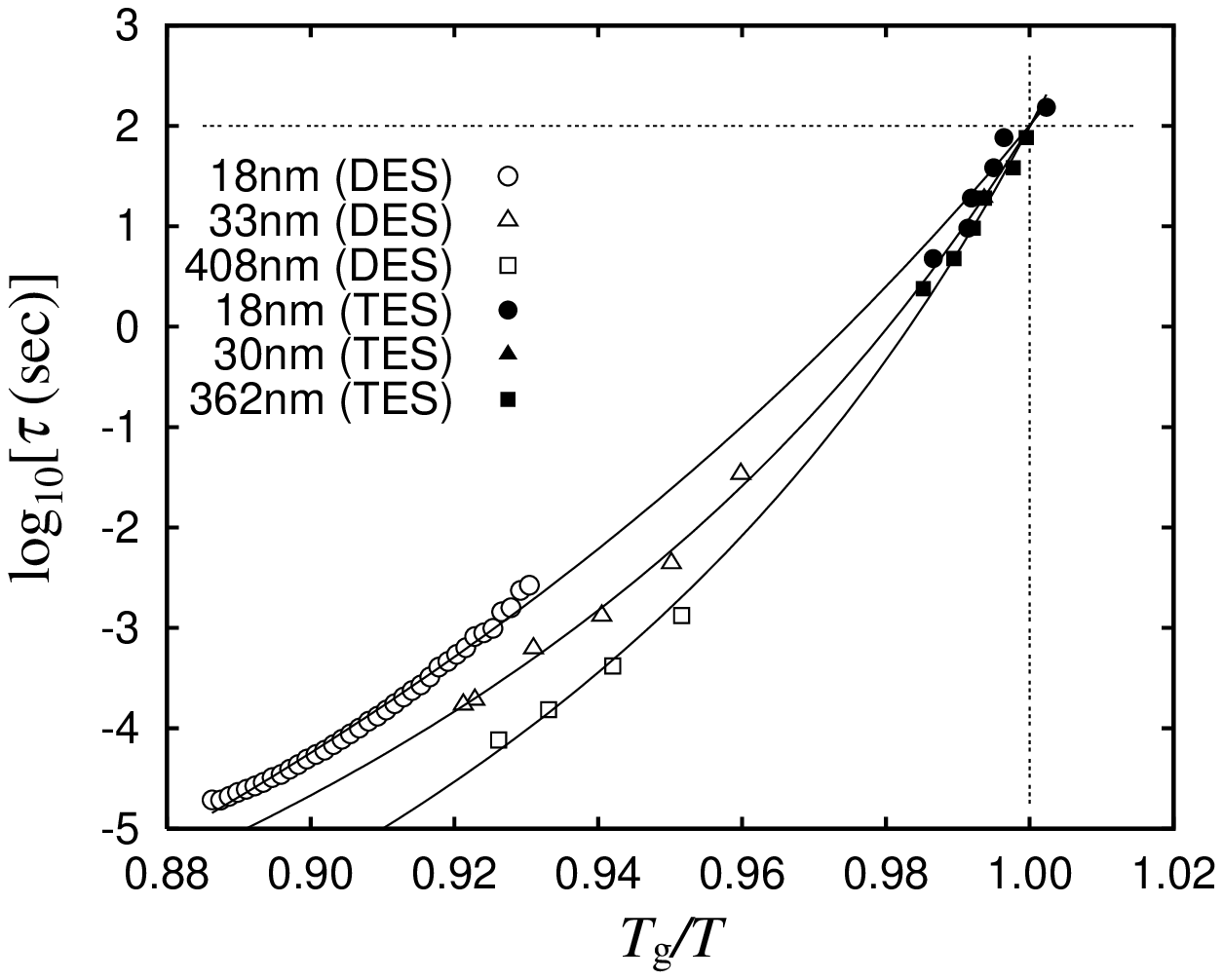}
\vspace{0.2cm}
{\setlength{\baselineskip}{9pt}
\small
{\bf Figure 5.}  Relaxation time as a function of $T_{\rm g}/T$ for
 various values of $d$. 
}

\vspace{0.8cm}
\end{minipage}

\vspace*{0.3cm}\noindent{\bf Acknowledgments}\\

\vspace{-0.2cm}\noindent 
This work was partly supported by a Grant-in-Aid from the Ministry 
of Education, Science, Sports and Culture of Japan.

\end{document}